\documentclass[preprint,showpacs,showkeys,amsmath,amssymb,aps,doublespace]{revtex4}
\usepackage[dvips]{graphicx}
\usepackage{setspace}
\usepackage{amsmath}
\usepackage{amsfonts}
\usepackage{amssymb,color}

\begin{document}

\title{Discontinuous phase transition in a dimer lattice gas}

\author{Ronald Dickman\footnote{email: dickman@fisica.ufmg.br}
}
\address{
Departamento de F\'{\i}sica and
National Institute of Science and Technology for Complex Systems,
ICEx, Universidade Federal de Minas Gerais,
C. P. 702, 30123-970 Belo Horizonte, Minas Gerais - Brazil
}

\date{\today}

\begin{abstract}

I study a dimer model on the square lattice with nearest-neighbor exclusion as the only
interaction.  Detailed simulations
using tomographic entropic sampling show that as the chemical potential is
varied, there is a strongly
discontinuous phase transition, at which the particle density jumps by about 18\%
of its maximum value, $1/4$.  The transition is accompanied by the onset of
orientational order, to an arrangement corresponding to the $\{1/2,0,1/2\}$ structure
identified by Phares et al. [Physica B {\bf 409}, 1096 (2011)]
in a dimer model with {\it finite} repulsion at {\it fixed} density.
Using finite-size scaling and Binder's cumulant, the expected scaling behavior
at a discontinuous transition is verified in detail.
The discontinuous transition can be understood qualitatively
given that the model possesses eight equivalent maximum-density configurations,
so that its coarse-grained description corresponds to that of the $q=8$ Potts
model.

\end{abstract}


\maketitle

\section{Introduction}

Lattice models of the fluid-solid phase transition have been studied
for many years \cite{runnels,gaunt-fisher,hall-stell,runnels72}.  Athermal models, characterized by
purely excluded volume interactions, are of particular interest in this
context for their simplicity, given that (off-lattice) systems of hard spheres
and hard disks exhibit fluid-solid transitions \cite{fernandes}.
Analyzing athermal lattice models of this kind is relevant not only to the
equilibrium melting transition, but to modeling diffusion in dense fluids,
glassy dynamics, disordered solids, and granular materials.

In the simplest lattice gas models, the fluid-solid transition is
continuous.  Thus the model with nearest-neighbor exclusion on
bipartite lattices belongs to the Ising universality class, while the
corresponding model on the triangular lattice - the hard-hexagon model -
has a transition in the $q=3$ Potts universality class \cite{baxterhex}.  Lattice gases with
extended exclusion regions on the square lattice were recently studied by
Fernandes et al. \cite{fernandes}.  For exclusion of up to $n$-th neighbors,
these authors found, for $n=1$ to 5, that the fluid-solid transition is
discontinuous for $n=3$ and 5, and continuous in the other cases.
What causes the transition to be continuous or not is as yet unclear.
It is therefore of interest to study further examples, with the goal of
identifying which features of a model determine the nature of the phase transition.

Considerable attention has also been given to lattice dimer models, in which each
molecule occupies a pair of adjacent sites.  In the two-dimensional case the model represents
adsorption of diatomic molecules on a crystalline surface, which has
motivated extensive study \cite{phares93,phares2011,pasinetti,ryzsko-binder06}.
(A complete review of this literature is
not feasible here; the interested reader will find further references in the papers cited.)
For purely on-site
excluded-volume interactions, it is known that no phase transition occurs \cite{wu09,heilmann-leib}.
Inclusion of nearest-neighbor or longer-range interactions, whether attractive or repulsive,
may lead to the appearance of new phases.  In addition to transitions between fluid phases
of high and low density (as in the monomer lattice gas), phases with periodic structure
may also arise.  Phase diagrams of lattice dimer systems have been studied via
transfer-matrix approaches \cite{phares93,phares2011} and Monte Carlo
simulation \cite{pasinetti,ryzsko-binder06}.

In this work I study a lattice gas composed of dimers that occupy neighboring
sites on the square lattice, with nearest-neighbor exclusion (see Fig. 1).
Using Monte Carlo simulations, I show that this model exhibits a
strongly discontinuous transition.  The high-density phase possesses orientational
order.  To obtain precise results on the phase diagram and thermodynamic
properties, I use tomographic entropic sampling \cite{tomographic}.  This Monte Carlo
simulation method has proved to be efficient and reliable in studies of continuous
phase transitions in spin models and lattice gases \cite{tomographic,isingafm}.
Thus a further motivation for the
present study is to assess the utility of tomographic sampling for discontinuous phase
transitions.  The results confirm that the method is indeed useful in this context:
a single set of simulations enables one to calculate the thermodynamic properties
(including the pressure and the absolute entropy), as continuous functions of the
chemical potential, without metastability effects.  Detailed comparison with theoretical
predictions for finite-size scaling at a discontinuous phase transition confirm the
reliability of the simulation method, and lend support to the idea that the nature of the
phase transition is in part determined by the symmetry of the close-packed state.

The balance of this paper is organized as follows. In the next
section I define the model, followed in Sec. III by a description of
the simulation method.
Sec. IV presents the simulation results, while Sec. V contains a discussion
of these results and possible directions for future investigation.
A mean-field analysis using the pair approximation is discussed in the Appendix.

\section{Model}

Consider a lattice gas consisting of dimers that occupy pairs of nearest-neighbor sites on the square lattice,
and exclude other dimers from occupying the six nearest neighbor sites.  An example of a
maximum density configuration is shown in Fig. 1.
Since the interaction between molecules is purely via excluded volume, all allowed configurations
have the same energy and the model is said to be {\it athermal}.
Let ${\cal C}$ denote an allowed configuration, i.e., a set of $N$ dimer positions that
respect the excluded volume condition.
In thermal equilibrium, in the grand canonical
ensemble, on a lattice of ${\cal N}$ sites, the probability of this configuration is
$P({\cal C}) = e^{\mu N({\cal C})}/\Xi$, where $\Xi(\mu,{\cal N})$ is the grand partition function,
and $\mu$ denotes the chemical potential divided by $k_B T$.  (From here on I refer to $\mu$
simply as the chemical potential.)
In this work I study the
model on $L \times L$ lattices with periodic boundaries.

The grand canonical partition function is given by

\begin{equation}
\Xi(\mu,L) = \sum_{N=0}^{N_{max}} z^N \Omega (N,L)
\end{equation}

\noindent where $z = e^\mu$ and $\Omega (N,L)$ is the number of distinct configurations of
$N$ molecules on a square lattice of $L \times L$ sites with periodic boundaries; the maximum
number of molecules is $N_{max} = L^2/4.$  The pressure is obtained directly from
$\pi = p/k_B T = \ln \Xi/L^2$, while the mean number of molecules per site is

\begin{equation}
\rho = \frac{\langle N \rangle}{L^2} = \frac{1}{\Xi} \sum_N N z^N \Omega (N,L)
\end{equation}

\noindent An analogous formula furnishes $\langle N^2 \rangle$, from which we can calculate
var$(N)/L^2$, which is proportional to the compressibility.  One may further define
an orientational order parameter,

\begin{equation}
m = \frac{1}{L^2 \Xi} \sum_N \langle |N_h - N_v| \rangle_N z^N \Omega (N,L),
\end{equation}

\noindent where $\langle |N_h - N_v| \rangle_N$ is the ``microcanonical" (fixed-$N$)
average of the absolute difference between the numbers of horizontally and vertically
oriented molecules.  The second and fourth moments of this quantity are also studied.

\begin{figure}[!htb]
\includegraphics[clip,angle=0,width=0.8\hsize]{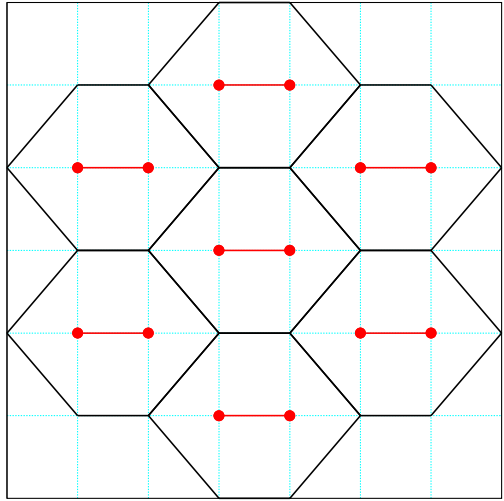}
\vspace{-4em}

\caption{A close-packed arrangement of dimers on the square lattice.
Since each monomer excludes occupancy at its nearest-neighbors, the arrangement forms
a hexagonal lattice.
}
\label{hex}
\end{figure}

The simulations reported in Sec. IV indicate that the transition between the low-density, unordered phase and
the high-density phase is discontinuous.  As discussed below, this can be understood on the
basis of symmetry considerations.
In particular, the fact that there are eight equivalent
close-packed configurations of the kind shown in Fig. 1 suggests that this model
is equivalent to the 8-state Potts model.
Another route to this conclusion arises from the observation
that the maximal density structure of Fig. 1 has been
identified as the first of a set of zero-entropy
``cusps" in a dimer model with strongly repulsive, but finite, nearest-neighbor interactions
\cite{phares93,phares2011}.  (These authors denote the structure of Fig. 1 as $\{1/2,0,1/2\}$.)
Varying the temperature at fixed density, $\rho = 1/4$,
a discontinuous phase transition is observed\cite{phares2011,pasinetti}.
The present model corresponds to the zero-temperature limit of the model studied
in \cite{phares93,phares2011} (i.e., nearest-neighbor pairs are prohibited).
Since the phase transition between ordered and disordered phases is
discontinuous at {\it finite} temperature, it should also be so at zero temperature (see Fig. 2).

\begin{figure}[!htb]
\includegraphics[clip,angle=0,width=0.8\hsize]{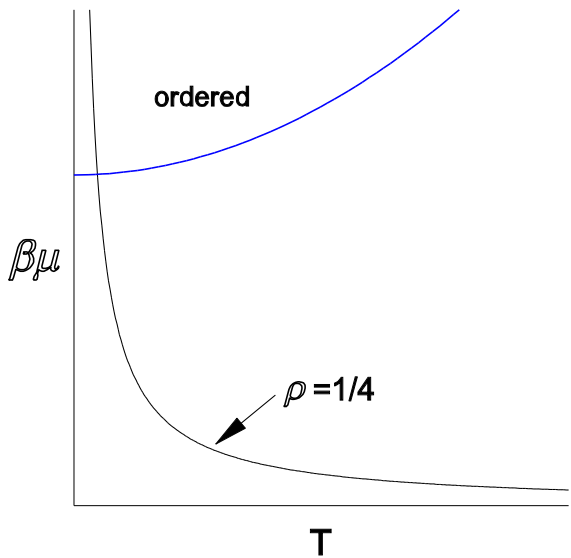}
\vspace{-4em}

\caption{Simplified (schematic) phase diagram of the model with {\it finite}
repulsion, in the temperature-$\beta \mu$ plane.  The studies of Refs. \cite{phares2011}
and \cite{pasinetti} are realized at density $\rho=1/4$; $\beta \mu$ tends to infinity
as $T \to 0$ along this line.  The present study corresponds to $T=0$.  Since the transition between
ordered and disordered phases is discontinuous along the line $\rho=1/4$, it should also be
along the line $T=0$.
}
\label{Tbetamu}
\end{figure}

Observe that the orientational order parameter is given
by $m = L^{-2} \partial \ln \Xi/\partial h$, where
$h$ is an external field coupled to the particle orientation; then $\chi = \partial m/\partial h$
is the orientational susceptibilty, proportional to var$[m]$.
Note however that $m$ is {\it not} the Potts-model order parameter, which involves the
fractions of dimers in each of the eight close-packed patterns.  (The latter quantity is not studied
in the present work.)  Since each close-packed
state exhibits perfect orientational ordering, it nevertheless seems reasonable to expect
$m$ to behave similarly to the Potts order parameter.

\section{Simulation Method}

As is evident from the foregoing discussion, one requires the configuration numbers
$\Omega (N,L)$ and the microcanonical averages of $|N_h - N_v|$ in order to evaluate
the quantities of interest.  These are obtained via
tomographic entropic sampling\cite{tomographic}, a Monte Carlo
method in which an initial estimate for the $\Omega (N,L)$ is refined in a series of iterations,
as in the well known Wang-Landau sampling (WLS) method \cite{wanglandau}.  As in WLS,
I use a target probability
distribution that is uniform on the set of energy values (or, as in the present case,
on the set of $N$ values.)
Different from WLS, the estimates (denoted $\hat{\Omega}(N,L)$) are updated only at the end of an
{\it iteration}, using

\begin{equation}
\hat{\Omega}_{n} (N,L) = \frac{H_n (N)}{\overline{H}_n} \hat{\Omega}_{n-1} (N,L)
\label{refine}
\end{equation}

\noindent where $H_n (N)$ is the histogram at iteration $n$, i.e., the number of visits
to configurations with exactly $N$ dimers, and $\overline{H}_n$ is the mean number of visits
per class.
Each iteration consists in ten rather long studies (each
involving $2 \times 10^7$ lattice updates) starting from diverse initial configurations.
(Three studies begin with a filled lattice, with all dimers oriented vertically,
three with all dimers oriented horizontally, and four starting from an empty lattice.)
The final estimates $\hat{\Omega}$ are obtained using 5-7 iterations of this kind.
The estimates of $\langle |N_h - N_v| \rangle_N$, $\langle (N_h - N_v)^2 \rangle_N$,
and $\langle (N_h - N_v)^4 \rangle_N$ are calculated at the final iteration.
Further details on the method may be found in \cite{tomographic}.

I study system sizes $L=8$, 16, 32, 64, and 128.  For the smallest
size, I begin with a uniform distribution, $\hat{\Omega}_0 (N) = 1, \; \forall N$; after the first
iteration the estimates, $\hat{\Omega}_1 (N)$, are quite close to their final values.
For $L=16$, the
initial estimate $\hat{\Omega}_0 (N)$ is obtained by interpolating the entropy density
$s(\rho) = \ln \hat{\Omega} (\rho L^2/2,L)/L^2$, using the final estimates in the
$L=8$ study, as explained in \cite{tomographic}.  The initial estimates for larger
systems are generated in an analogous manner.

As noted above, the target probability distribution is uniform on the set of
$N$ values, 0, 1, 2,...,$N_{max}$.  This means that the probability of
configuration ${\cal C}$, having $N$ dimers, satisfies $P({\cal C}) \propto 1/\Omega(N,L)$.
Using a Metropolis-like procedure,
the probability $P[{\cal C} \to {\cal C}']$ for a transition from configuration ${\cal C}$
to a trial configuration, ${\cal C}'$, should satisfy,

\begin{equation}
\frac{P[{\cal C} \to {\cal C}']} {P[{\cal C}' \to {\cal C}]}= \frac{\Omega(N,L)}{\Omega(N',L)},
\label{trprob}
\end{equation}

\noindent where $N$ and $N'$ are the numbers of dimers in the current and trial configurations,
respectively.
In this way, the transition probabilities satisfy detailed balance with respect to the target
distribution.
Since the $\Omega(N,L)$ are unknown, the transition probabilities are evaluated using the
current estimates, $\hat{\Omega}$.

The simulation algorithm uses an insertion/removal dynamics.  In a removal step, one of the $N$
existing dimers is selected at random and deleted.  In an insertion step, one of the $B$ lattice
bonds currently available for insertion is selected and a dimer placed there.  (A list of
available bonds is maintained.)  The probabilities for insertion and removal are as follows.
Let ${\cal C}$ be a configuration with $N$ dimers and $B$ available bonds.
Starting from this configuration, insertion is chosen with probability

\begin{equation}
p_{ins} = \frac{1}{2} \frac{B \Omega(N,L)}{B \Omega(N,L) + (N+1) \Omega(N+1,L)},
\label{pins}
\end{equation}

\noindent while removal is chosen with probability 1/2 (independent of the configuration).
Unlike insertion, however, removal is not always accepted.  Suppose the insertion
contemplated above results in a configuration ${\cal C}'$.  To satisfy the detailed balance
relation, Eq.  (\ref{trprob}), a removal step
yielding configuration ${\cal C}$ should be accepted with probability

\begin{equation}
p_{acc} = \frac{(N+1) \Omega(N+1,L)}{(N+1) \Omega(N+1,L) + B \Omega(N,L)}
\label{pacc}
\end{equation}

\noindent This is because the specific transition ${\cal  C} \to {\cal C}'$ then
occurs with probability $p_{ins}/B$ while the probability of the inverse transition
is $p_{acc}/[2(N+1)]$.  In general, the acceptance probability for removal, starting from
a configuration with $N$ dimers, is

\begin{equation}
p_{acc}(N,B') = \frac{N \Omega(N,L)}{N \Omega(N) + B' \Omega(N-1,L)}
\end{equation}

\noindent where $B'$ is the number of available bonds in the {\it target} configuration.
Maintaining of a list of available bonds and calculating the change in the
number of such bonds before accepting a removal involves some computational overhead,
but this is more than compensated by the improved efficiency, as compared with a
``blind" insertion procedure.  To choose the next transition, a random number $z$,
uniformly distributed on $[0,1)$ is generated, and if $z < p_{ins}$ insertion is performed,
while if $z > 1/2$ removal is attempted.  In case $p_{ins} < z < 1/2$, no transition is
performed, and the current configuration is counted once again in the histogram and the
microcanonical averages.

For each system size, we have the exact values $\Omega(N_{max},L) = 8$, $\Omega(0,L)=1$,
$\Omega(1,L) = 2L^2$, and $\Omega(2,L) = L^2 (2 L^2-23)$.  I use the latter value to
fix the absolute entropy, via

\begin{equation}
\hat{\Omega}_{abs} (N,L) = L^2 (2 L^2 -23) \frac{\hat{\Omega}(N,L)}{\hat{\Omega}(2,L)}.
\label{omegaabs}
\end{equation}

\noindent This is required since the simulations only furnish ratios between configuration numbers.

\section{Results}

I turn now to results for thermodynamic properties.  Figure 3 shows the particle density $\rho$
as a function of chemical potential $\mu$, while Fig.~4 shows the pressure versus density, calculated
directly from the grand canonical partition function.  The hallmarks of a discontinuous phase transition
are immediately evident: a sudden change in density with varying chemical potential,
associated with a nearly flat region in the isotherm
$\pi (\rho)$.  The orientational order
parameter $m$ also appears to develop a discontinuity with increasing system size, as shown in
Fig. 5.  From this graph it appears that, in the thermodynamic limit, the low-density phase
exhibits no orientational order, while in the high-density phase nearly all molecules have the
same orientation.  A striking aspect of these graphs is the absence of metastability or hysteresis
effects commonly observed in conventional Monte Carlo simulations of a system exhibiting a
discontinuous phase transition.

\begin{figure}[!htb]
\includegraphics[clip,angle=0,width=0.8\hsize]{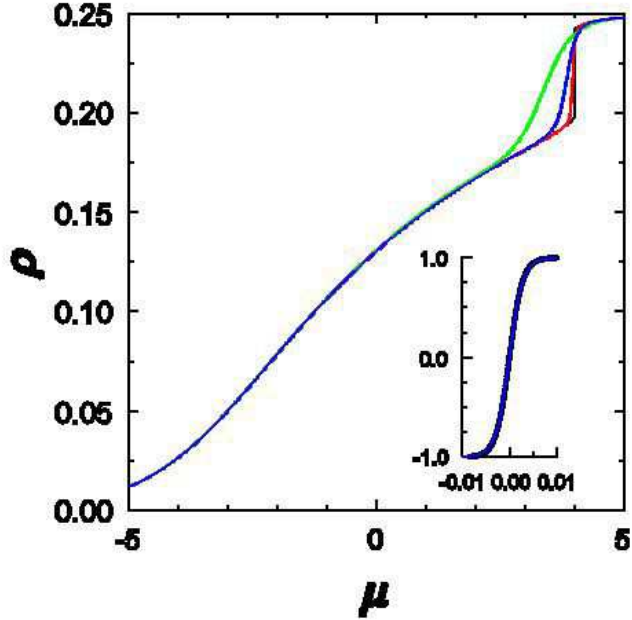}
\vspace{-2em}

\caption{Particle density $\rho$ versus chemical potential $\mu$ for
system sizes $L=16$, 32, 64, and 128, in order of increasing steepness.  Uncertainties
are smaller than the line thickness.  Inset: data for $L=128$ plotted as $\tilde{\rho}=
2(\rho - \overline{\rho})/\Delta \rho$ versus $\tilde{\mu} = \mu - \mu^*_L$, compared with
the scaling function $\tilde{\rho} = \tanh b \tilde{\mu}$.  Here $\overline{\rho}$ is the
mean mean of the coexisting densities and $\Delta \rho$ their difference; the constant
$b = 347.2$.
}
\label{rvmb}
\end{figure}

\begin{figure}[!htb]
\includegraphics[clip,angle=0,width=0.8\hsize]{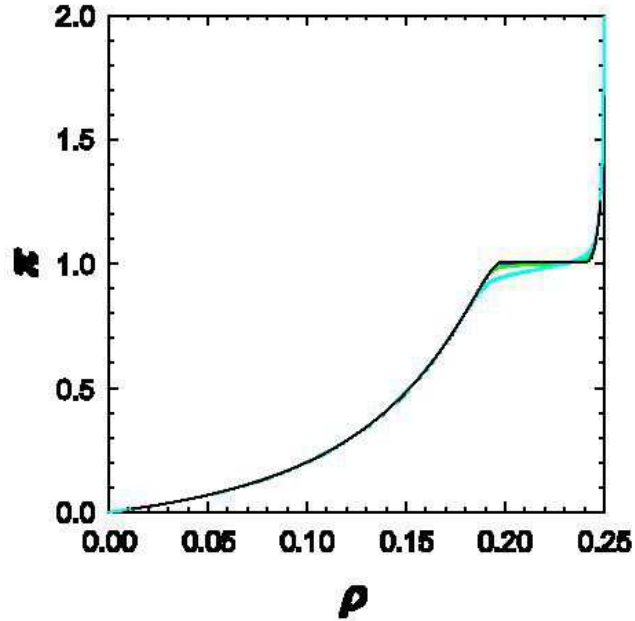}
\vspace{-2em}

\caption{Pressure $\pi$ versus particle density $\rho$ for
system sizes $L=16$, 32, 64, and 128, in order of increasing flatness in the
transition region.  Relative uncertainties in $\pi$ are of order $10^{-6}$.
}
\label{prdm}
\end{figure}

\begin{figure}[!htb]
\includegraphics[clip,angle=0,width=0.8\hsize]{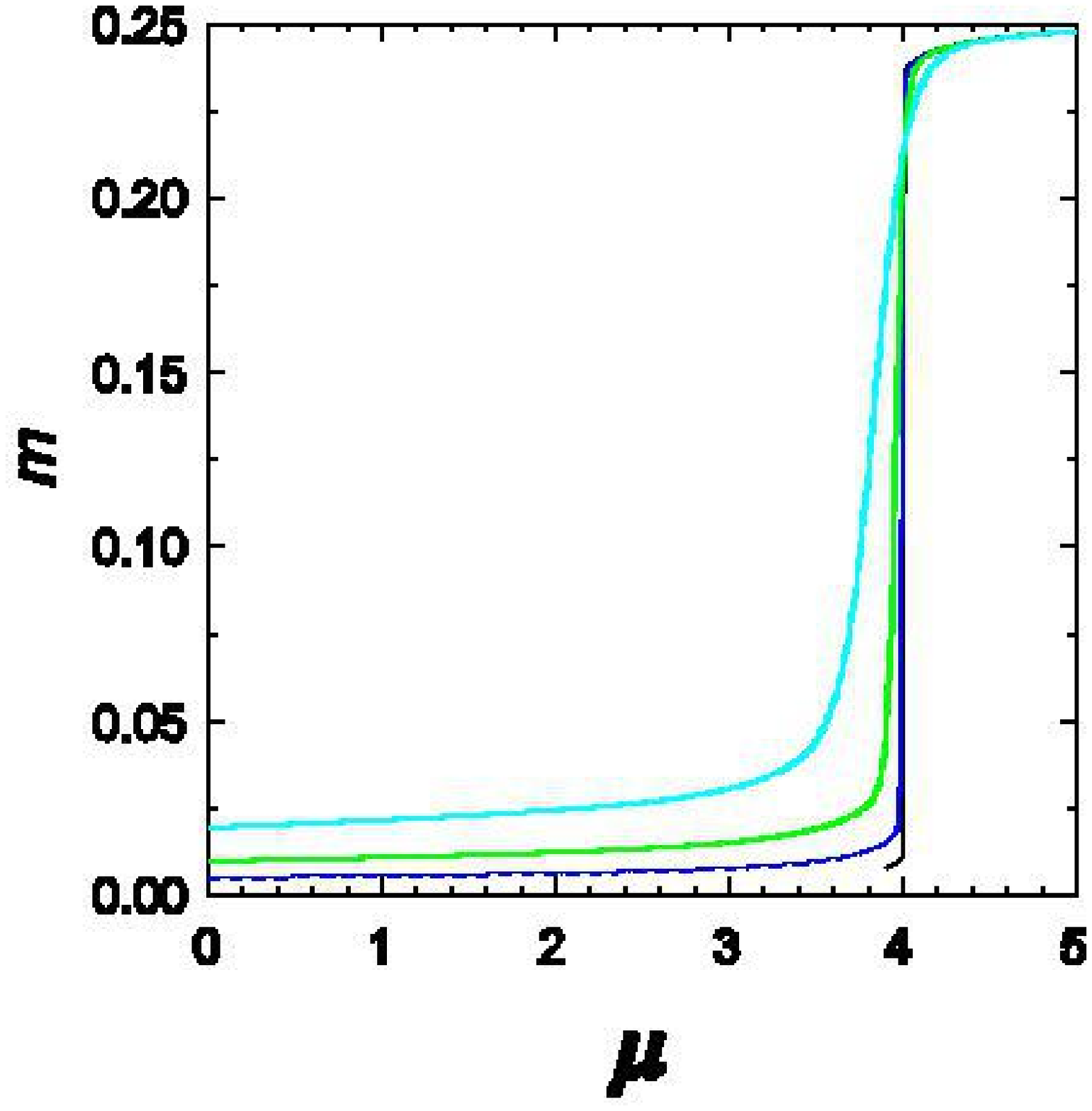}
\vspace{-2em}

\caption{Orientational order parameter $m$ versus chemical potential
$\mu$ for
system sizes $L=16$, 32, 64, and 128, in order of increasing steepness in the
transition region.  Uncertainties
are smaller than the line thickness.
}
\label{mvmu}
\end{figure}

To pinpoint the transition, and elucidate its nature, it is helpful to review some results
on finite-size scaling at discontinuous phase transitions \cite{binder-landau,lee-kosterlitz,BK90,BKMS}.
Consider a system that (in the thermodynamic limit) suffers a discontinuous phase transition
at temperature $T^*$.  Let $T^*_{L,c}$ denote
the temperature at which the specific heat takes it maximum value, in a system of linear size $L$.
The results of Borgs and Koteck\'y \cite{BK90} suggest that at a discontinuous phase transition
at which only two phases coexist, the shift, $\Delta T^*_{L,c} = |T^* - T^*_{L,c}|$, decays $\sim L^{-2d}$
in $d$ dimensions, for large $L$.
In the case of a discontinuous transition in a $q$-state Potts model, however,
the shift decays more slowly\cite{BKMS}: $\Delta T^*_{L,c} \sim L^{-d}$;
the width of the transition region also decreases $\sim L^{-d}$.
For periodic boundary conditions,
the energy and specific heat per site follow the scaling forms
$e \sim (e_1+e_2)/2 - (1/2)(e_1-e_2) \tanh [(e_1 - e_2)(\beta - \beta^*)L^d/2 + \ln q/2]$ and
$c \sim L^d k_B \beta^2 / \cosh^2 [(e_1 - e_2)(\beta - \beta^*)L^d/2 + \ln q/2]$.

In the present instance, the chemical potential $\mu$ is the temperature-like variable, with
the particle density $\rho$ playing the role of the energy and $\kappa \equiv$ var$(\rho)$ that of the
specific heat. I denote the the chemical potential at which $\kappa$ takes its maximum by
$\mu^*_{L,\kappa}$; that at which the variance, $\chi$ of the orientational order parameter
takes its maximum is denoted by $\mu^*_{L,\chi}$.  Based on the results of \cite{lee-kosterlitz,BKMS},
one expects the shifts, $\Delta \mu^*_L = |\mu^* - \mu^*_L|$, in chemical potential
to decrease $\sim L^{-2}$, and for $\rho$ and
var$(\rho)/L^2$ to be governed by scaling functions $\sim \tanh \overline{\mu}$
and $\cosh^{-2} \overline{\mu}$, respectively,
where the scaling variable $\overline{\mu} \propto L^2 (\mu - \mu^*_L)$.  The inset of Fig. 3
confirms the hyperbolic tangent form of the density in the region of the transition.

As expected,
the variances of $\rho$ and $m$ exhibit sharp maxima in the vicinity of the transition; the associated
chemical potential values are plotted versus $L^{-2}$ in Fig. 6.
Quadratic extrapolation (versus $L^{-2}$) of the data for
$L \geq 16$ yields the transition values $\mu^* = 4.0076(1)$
(from var$[\rho]$) and 4.0077(1) (from var$[m]$) leading to the estimate $\mu^* = 4.00765(15)$.

\begin{figure}[!htb]
\includegraphics[clip,angle=0,width=0.8\hsize]{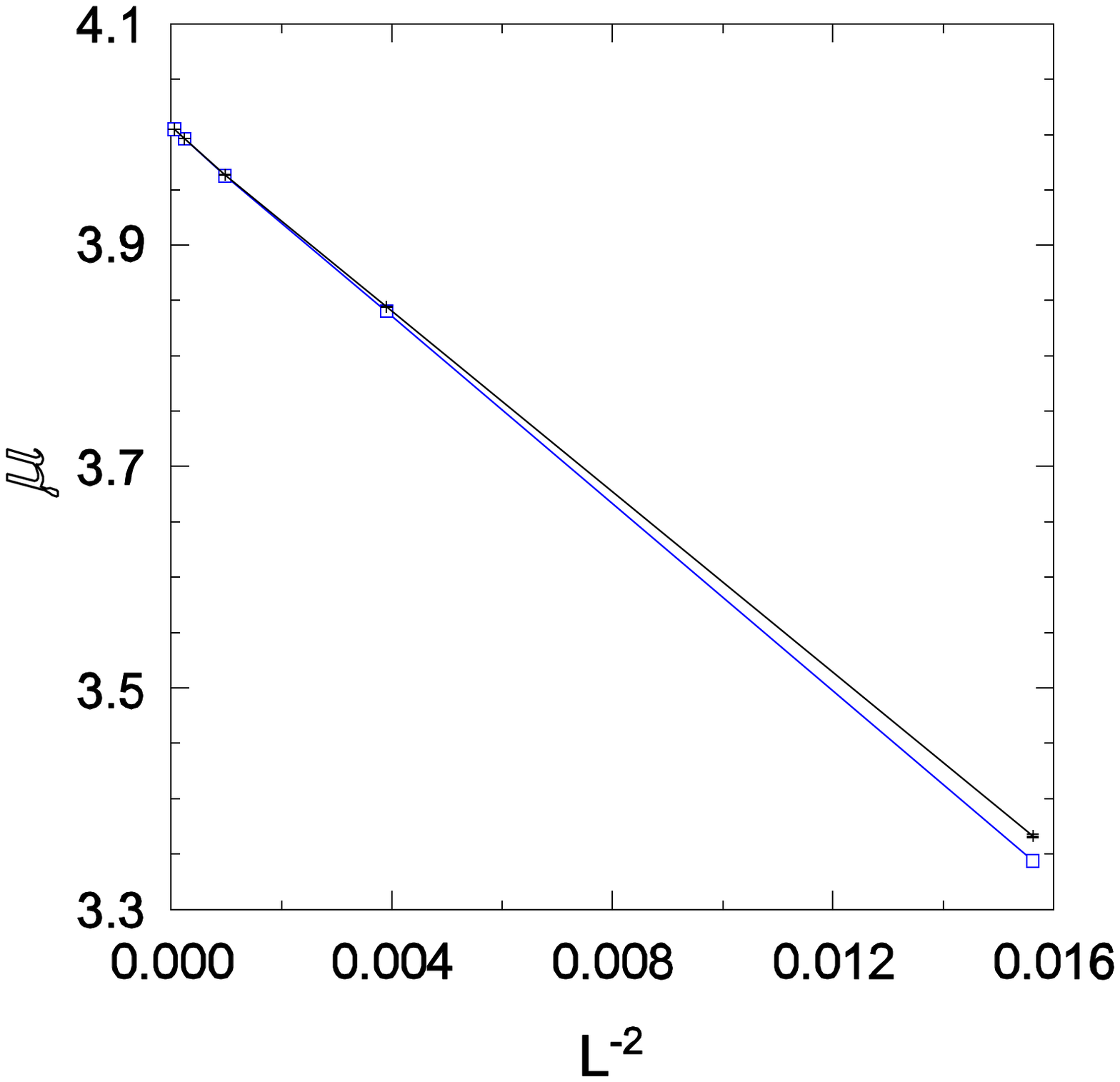}
\vspace{-2em}

\caption{Chemical potentials at which var$[\rho]$ (upper) and var$[m]$ (lower) attain
their maxima versus $1/L^2$.  Error bars
are smaller than the symbols.
}
\label{mustar}
\end{figure}

To estimate the densities of the fluid and solid phases at coexistence, I analyze the
data for $\rho (\mu)$ on an interval near $\mu^*$, but far enough away that the results
for the two largest system sizes are equal, to avoid effects of finite-size rounding.
(The specific ranges are $3.5 \leq \mu \leq 3.96$
and $4.03 \leq \mu \leq 4.5$.)  Extrapolating these data to $\mu^*$ using a quartic
polynomial yields coexisting
densities $\rho_f = 0.1974(1)$ and $\rho_s = 0.2421(1)$.  Applying the same procedure to the
pressure data, I find $\pi=1.00748(2)$ using the fluid-phase data, and
$\pi = 1.00760(2)$ from the solid phase;  I therefore estimate the
pressure at coexistence as $\pi_{coex} = 1.00754(10)$.
Figure 7 shows a typical configuration in the coexistence region $(\rho = 0.22$);
fluid- and solidlike regions are clearly visible.

\begin{figure}[!htb]
\includegraphics[clip,angle=0,width=0.8\hsize]{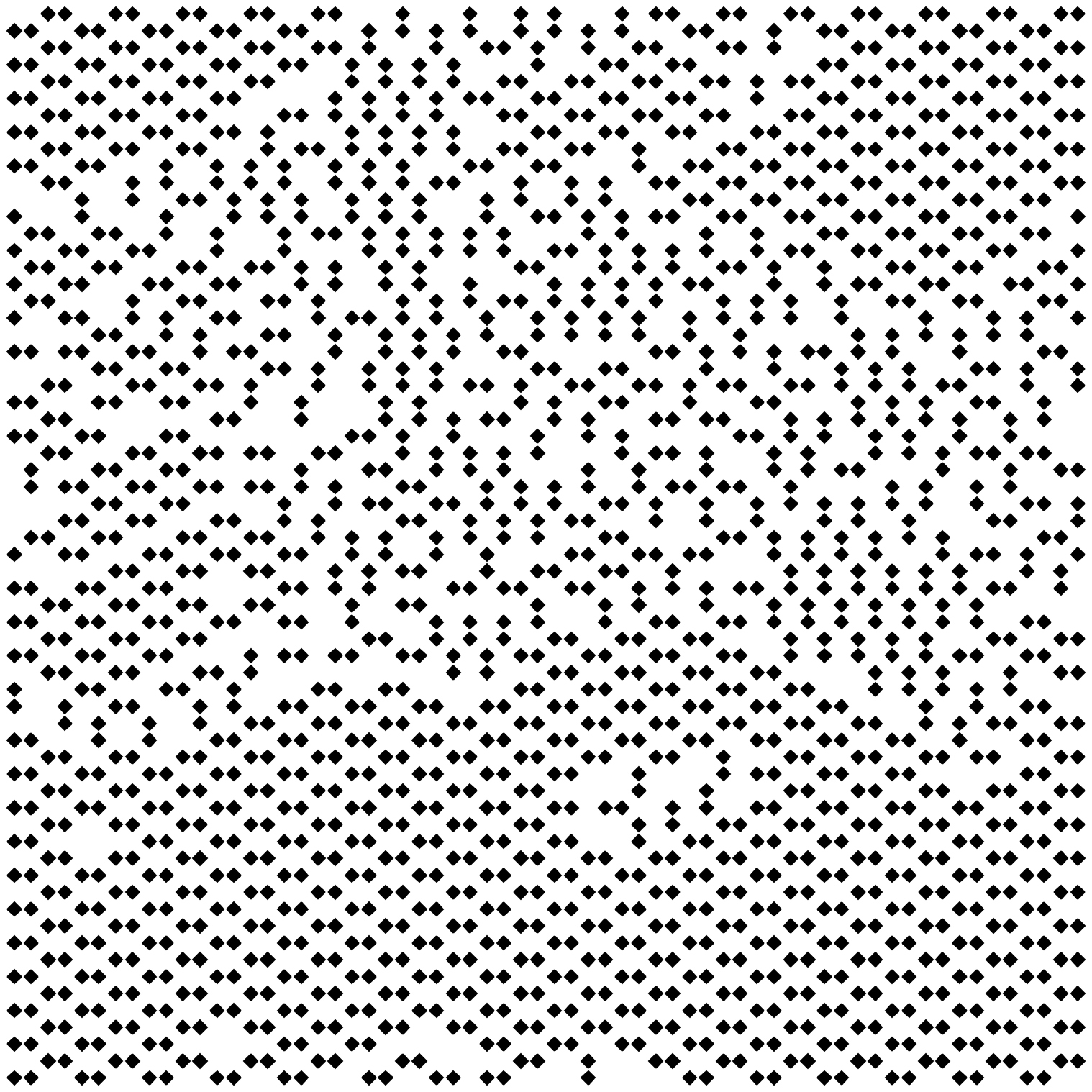}
\vspace{-2em}

\caption{A typical configuration in the coexistence region, $\rho=0.22$, $L=64$.
}
\label{rho22}
\end{figure}

Having determined the coexistence parameters, I return to the scaling properties at the
transition.  Figure 8 confirms that
$\overline{\kappa} \equiv \kappa/L^d \propto 1/\cosh^2 [L^d (\mu-\mu^*_L)]$.
In the case of $\overline{\chi} \equiv \chi/L^2$
(Fig. 9), there are significant deviations from the expected scaling, with the
maximum of $\chi$ apparently growing faster than $L^2$.  Analysis of the data for $L = 16$ - 128
yields $\chi_{max} \sim L^{2.18(1)}$, with some suggestion of a smaller exponent at the
largest system size.  This may be associated with the fact that the jump in $m$ grows with
increasing $L$, as the value of $m$ in the fluid phase slowly approaches zero as $L$ grows.
It thus appears likely that larger systems would be required to observe the scaling of $\chi$.
In the case of the pressure, the deviation from the coexistence value, $\Delta p = p-p_{coex}$,
appears to decrease $\propto 1/L^2$ in the coexistence region, as shown in Fig. 10.

\begin{figure}[!htb]
\includegraphics[clip,angle=0,width=0.8\hsize]{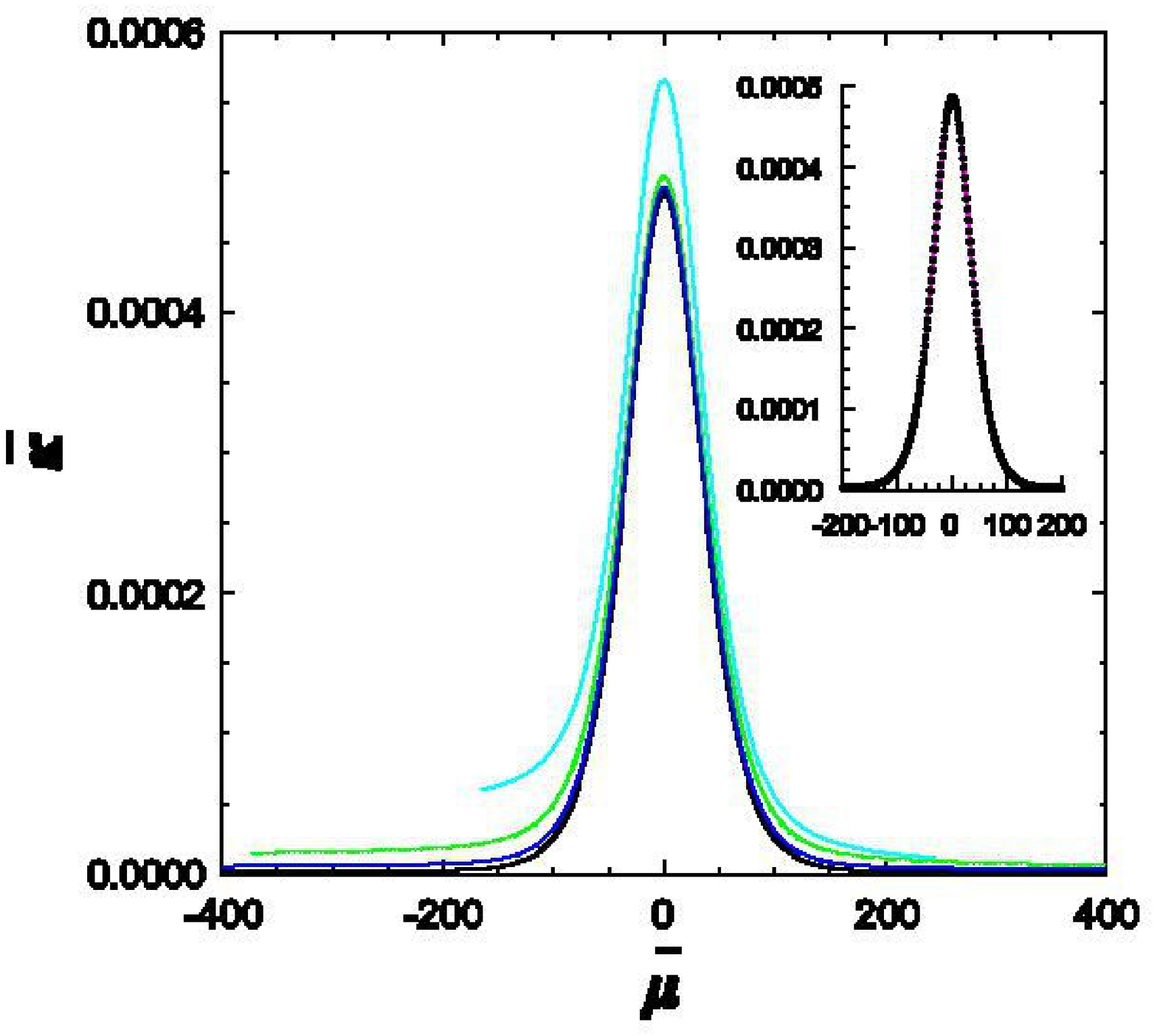}
\vspace{-1em}

\caption{Scaling plot: $\overline{\kappa} \equiv \kappa/L^2$ versus
$\overline{\mu} \equiv L^2 (\mu-\mu^*_L)$ for (upper to lower) $L=16$, 32, 64, and 128.
Relative uncertainties in $\overline{\kappa}$ are $< 4\%$.  Inset: data for $L=128$ (points)
compared with scaling function $\overline{\kappa} = A /\cosh^2 b \overline{\mu}$, using
$A = 4.87 \times 10^{-4}$ and $b = 0.02182$.
}
\label{varrho1}
\end{figure}

\begin{figure}[!htb]
\includegraphics[clip,angle=0,width=0.8\hsize]{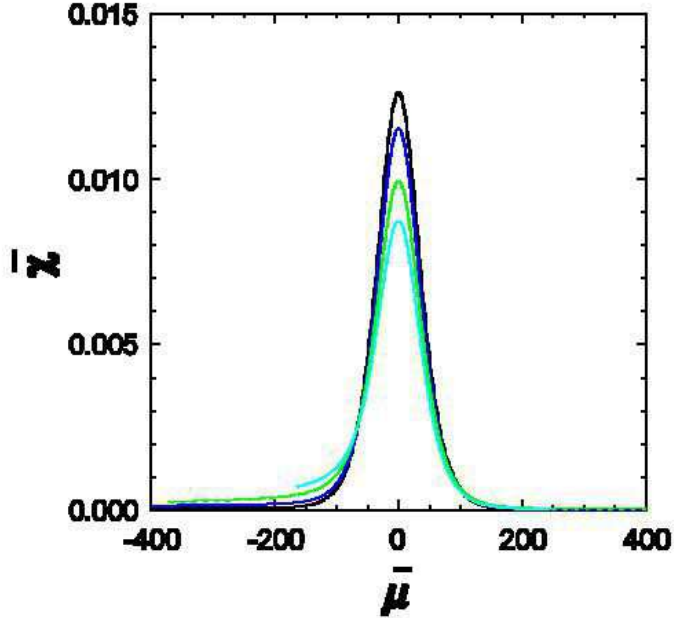}
\vspace{-2em}

\caption{Scaling plot: $\overline{\chi} \equiv \chi/L^2$ versus
$\overline{\mu} \equiv L^2 (\mu-\mu^*_L)$ for (lower to upper) $L=16$, 32, 64, and 128.
Relative uncertainties in $\overline{\chi}$ are $< 4\%$.
}
\label{varmsc}
\end{figure}

\begin{figure}[!htb]
\includegraphics[clip,angle=0,width=0.8\hsize]{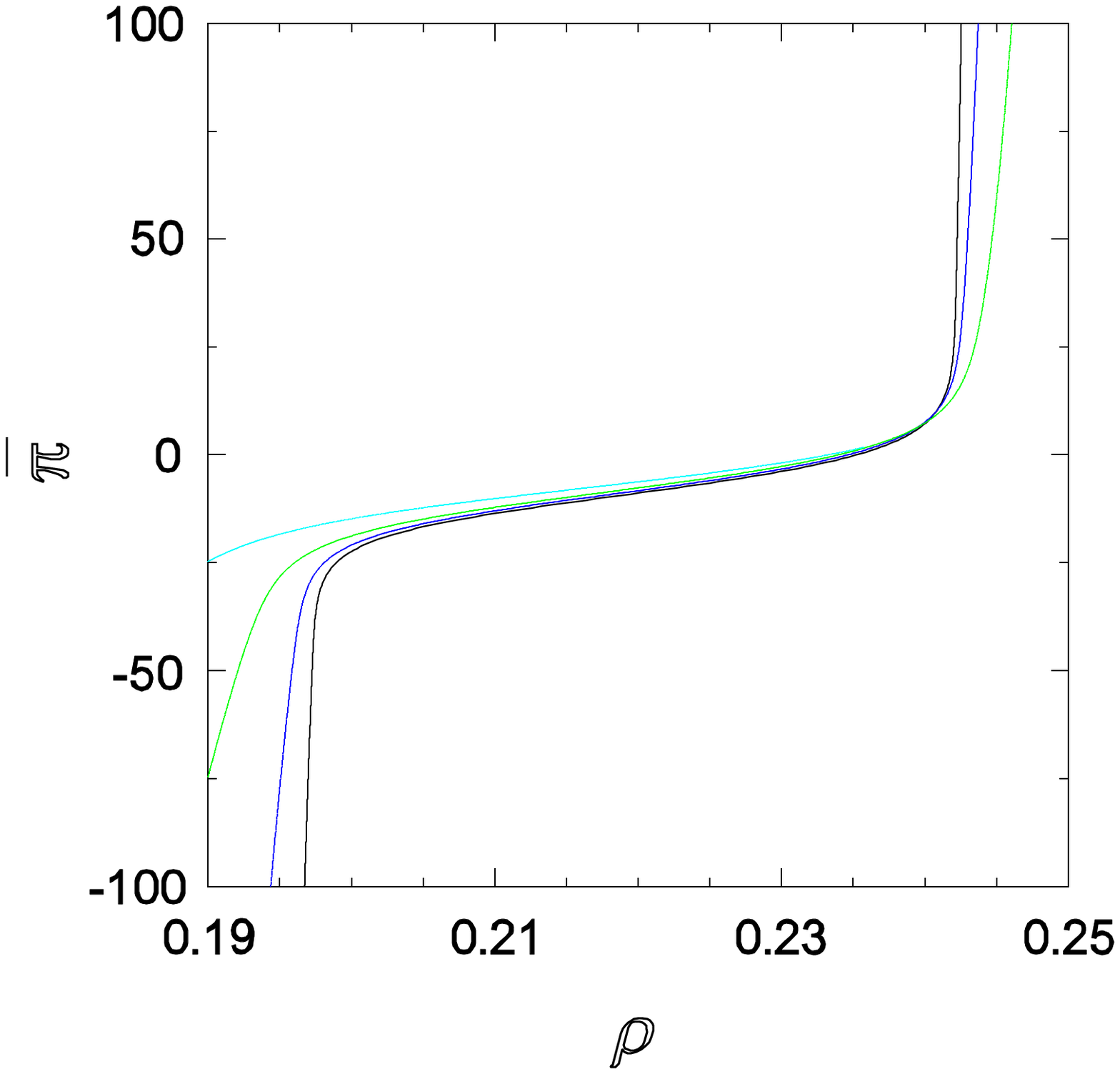}
\vspace{-2em}

\caption{Scaled excess pressure $\overline{\pi} \equiv L^2 (\pi - \pi_{coex})$
versus density for (lower to upper, on left) $L=16$, 32, 64, and 128.
}
\label{dpsc}
\end{figure}

Moment ratios such as Binder's cumulant \cite{binder81} are valuable in analysis of discontinuous
as well as continuous phase transitions.  At a (temperature-driven) discontinuous transition,
Binder's cumulant,
$V_4 \equiv 1 - \langle E^4 \rangle/3 \langle E^2 \rangle^2$ exhibits a minimum
given by\cite{BKMS,lee-kosterlitz}

\begin{equation}
V_{4,min} = \frac{2}{3} - \frac{1}{12} \left( \frac{e_1}{e_2} - \frac{e_2}{e_1} \right)^2
+ \frac{A^{(2)}}{L^d}
\label{v4min}
\end{equation}

\noindent where $A^{(2)}$ is a constant and $e_1$ and $e_2$ are the bulk energy densities
of the coexisting phases.  Identifying, as above, $\mu$ as the temperature-like
variable, and the density as the variable corresponding to the energy density,
I define the reduced cumulant as
$V_4 \equiv 1 - \langle N^4 \rangle/3 \langle N^2 \rangle^2$; Eq. (\ref{v4min}) is
expected to hold if we replace $e_i$ with $\rho_i$.  This is indeed borne out by the
simulation data: I find (for $L=16$ - 128) $V_4 = 0.6531(1) - 0.70(4)/L^2$.
Using Eq. (\ref{v4min}) with the estimates for the coexisting densities cited above,
I find lim$_{L \to \infty} V_4= 0.6526(1)$, quite close to the simulation result.
The chemical potential values at which $V_4$ takes its minimum extrapolate to $\mu = 4.0076(1)$,
consistent with the estimate for the transition point obtained previously.

The results discussed above support, in considerable detail, the conclusion that the
model exhibits a discontinuous phase transition.  Equivalence to the $q=8$ Potts model is
supported by the following observation.  The amplitude of the leading correction to the finite size shift,
i.e., the amplitude ${\cal A}$ in the relation $\Delta T^*_{L,c} \simeq {\cal A} L^{-d}$,
is given by\cite{BKMS} ${\cal A} = (\ln q)/(e_1 - e_2)$.  Identifying, as before, temperature
with chemical potential, and energy density with $\rho$, we may write, for the effective
number of states, $q_{eff} = \exp[{\cal A} (\rho_1 - \rho_2)]$.  The amplitude is
estimated from the data for $\mu^*_L$ by extrapolating the local slopes,
${\cal A}_L \equiv (\Delta \mu^*_L - \Delta \mu^*_{L/2})/[L^{-2} - (L/2)^{-2}]$ to $L \to \infty$.
The resulting estimate of ${\cal A} = 48.8(6)$ corresponds to $q_{eff} = 8.0(3)$, consistent with the
value expected.

Finally, it is of interest to note that a discontinuous phase transition is associated with a
nonconcave portion of the
density of states (or microcanonical entropy per site) $s(\rho) = \ln \Omega (\rho L^2,L)/L^2$.
(I employ units such that $k_B=1$.)  The entropy is in general quite difficult to
obtain using conventional simulation methods, but is accessible via entropic simulation.
The density of states is compared with the thermodynamic entropy per site,

\begin{equation}
s_{th} = \frac{1}{L^2} \sum_{\cal C} p({\cal C}) \ln p({\cal C})
       = \frac{1}{L^2} \sum_N \Omega(N,L) z^N - \rho \mu,
\label{sth}
\end{equation}

\noindent in Fig. 11.  At the scale of the main graph the microcanonical and thermodynamic
entropies are indistinguishable, but a detail of $s^* \equiv s + \mu^* \rho$ shows that
the former has a positive curvature in the transition region, while the latter, as expected,
is concave.

\begin{figure}[!htb]
\includegraphics[clip,angle=0,width=0.8\hsize]{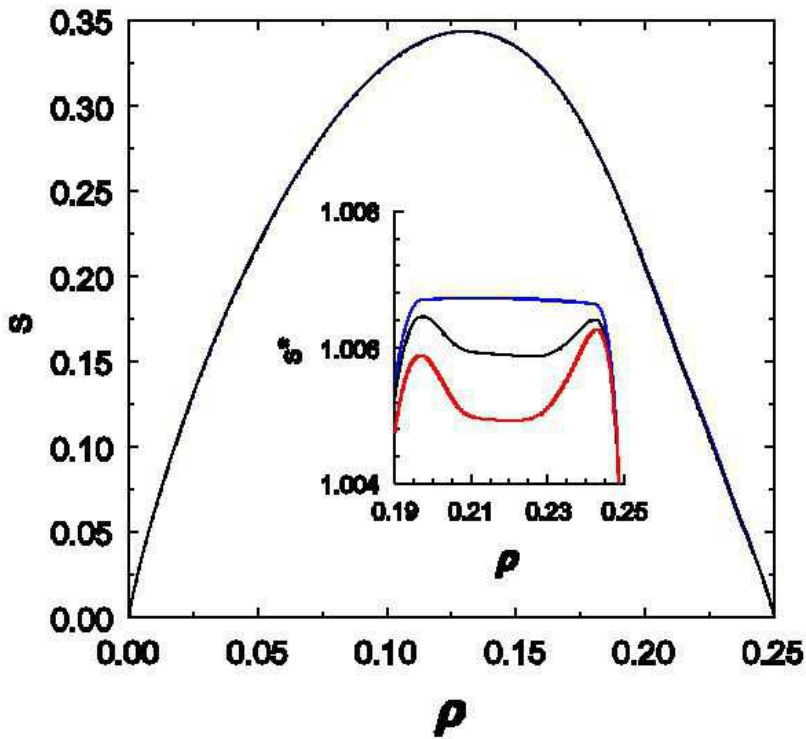}
\caption{Main graph: microcanonical and thermodynamic entropies
versus density for $L=128$; the two functions are indistinguishable at this scale.
Inset: $s^* = s + \mu^* \rho$ for (lower to upper) the microcanonical entropy ($L=64$
and 128), and the thermodynamic entropy.
}
\label{s128}
\end{figure}

\section{Discussion}

A dimer lattice gas with nearest-neighbor
exclusion on the square lattice exhibits a strongly discontinuous fluid-solid
transition accompanied by orientational ordering.  Tomographic entropic
sampling in the grand canonical ensemble proves to be well suited to the study
of thermodynamic properties, including the absolute entropy and the pressure.
The expected scaling properties of the Binder cumulant, compressibility and other properties
at the transition are confirmed, with the possible exception of the variance
of the orientational order parameter, for which larger system sizes need to be studied.
Further analysis of this model (and its counterparts on other two- and three-dimensional lattices),
should be of considerable interest, both as regards equilibrium
properties, metastability, particle diffusion, and nucleation dynamics, as well as nonequilibrium
aspects such as a possible glassy phase and behavior under an external drive.  Adding an
attractive interparticle potential, possibly with orientational dependence, one should expect a
richer phase diagram, including a liquid-vapor phase transition, nematic ordering, and
structural transitions in the solid phase.

A detailed theoretical analysis of the model appears
not to have been performed, although, as mentioned in Sec. II, there is strong reason to expect a
discontinuous transition given the results on the related model with finite
repulsion \cite{phares2011,pasinetti}.  The transfer-matrix approach of Refs. \cite{phares93,phares2011}
appears most suitable for this analysis.  In the Appendix, I show that a relatively simple
dynamic pair approximation yields a discontinuous phase transition, with coexisting densities
that are comparable to those found via simulation.

I close with a few general observations on this and related models.  It is worth recalling
that athermal lattice dimers with on-site exclusion only
(the {\it monomer-dimer} model, with monomers corresponding
to sites not occupied by dimers), do not exhibit a
phase transition \cite{wu09,heilmann-leib}.
The present study shows that
extending the exclusion to first neighbors, a discontinuous transition arises, in
contrast to the simple
(monomer) lattice gas, in which exclusion out to third neighbors is required for a discontinuous fluid-solid
transition.  The simple lattice gas with {\it nearest-neighbor} exclusion
exhibits a continuous transition, in the Ising universality class, as can be understood
on the grounds of symmetry, since there are two symmetric maximally occupied states.  For hard hexagons there
are three equivalent maximally occupied states, and the transition, which is again continuous,
belongs to the $q=3$ Potts model class \cite{baxterhex}.  For the dimer model studied here, there are eight
maximally occupied states, suggesting equivalence (under a suitable coarse-graining procedure)
to the $q=8$ Potts model, which exhibits a discontinuous phase transition.
This correspondence is in fact supported by the simulation results.

It is interesting to
note, in this context, that the hard-core lattice gas with exclusion out to third neighbors
possesses ten distinct maximally occupied configurations.  This model also suffers a discontinuous phase
transition \cite{fernandes}, as would be expected from its similarity to the $q=10$ Potts model.
It seems natural to conjecture, then, that a lattice gas having exactly $q$ equivalent
maximum-density states (or minimum-energy states) will exhibit a discontinuous phase transition
if $q$ is sufficiently large, i.e., $q > 4$ in two dimensions.  The converse to this statement does not,
however, appear to hold.  Thus, the number of maximum-density configurations of the lattice gas
with exclusion out to fifth neighbors is {\it infinite}, but the model nevertheless exhibits
a discontinuous transition \cite{fernandes}.

\begin{figure}[!htb]
\includegraphics[clip,angle=0,width=0.8\hsize]{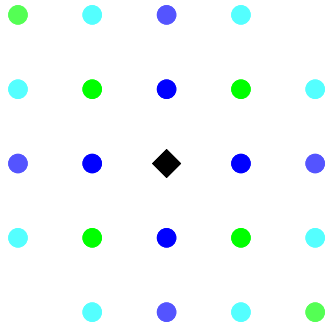}
\vspace{-4em}

\caption{Mapping to a particle model: exclusion zone of a particle (central diamond) on
sublattice A, corresponding to a dimer occupying a horizontal bond on the
original lattice.
}
\label{exzone}
\end{figure}

Finally, I note that the dimer model may also be mapped to a particle model, i.e., to a
lattice in which each bond of the original square lattice
corresponds to a {\it site} on the new lattice, which is also square.  Under this mapping, an occupied
horizontal bond in the original model corresponds to an occupied site in sublattice A of the
new lattice, while a vertical bond corresponds to a site in sublattice B.  The set of 22 excluded
sites (in the new lattice) includes all sites out to fourth neighbors of the central site, as well
as two fifth neighbors in opposite corners (see Fig.~12); the other two fifth neighbors
are not excluded.  The exclusion zone of a
particle in sublattice B is the same, except that the fifth neighbors not excluded by a particle in
sublattice A are excluded, and vice-versa.  The lattice gas with exclusion out to fourth neighbors
has a continuous fluid-solid transition, while exclusion out to fifth neighbors leads to a discontinuous
transition \cite{fernandes}.  The dimer model corresponds to an intermediate case,
with different orientations of the exclusion zone
associated with different sublattices.

\vspace{1cm}

\noindent{\bf Acknowledgments}
\vspace{1em}

I thank Prof. A. Ramirez-Pastor for communicating his results prior to publication.
This work was supported by CNPq, Brazil.
\vspace{1em}

\noindent{\bf Appendix}
\vspace{1em}

I consider
a two-site (pair approximation) mean-field analysis \cite{marro}.  This approach seeks the equilibrium
state of a dynamics in which dimers are inserted at available bonds at rate $z$, and
removed at unit rate.  To begin, consider a simplified a simplified description.
The probability that a dimer can be inserted at a given bond
is the probability that this bond as well as the six neighboring bonds be empty.
In the pair approximation this probability is estimated as

\begin{equation}
p_{free} = q \left(\frac{q}{y}\right)^6
\label{pfree}
\end{equation}

\noindent where $q$ denotes the fraction of unoccupied bonds and $y$ the fraction of unoccupied sites.
Let $p = N/2L^2$ denote the fraction of bonds occupied by dimers.  Noting that each dimer blocks
seven bonds, we may write $q = 1 - 7p$.  Each dimer occupies two sites, so the fraction of
vacant sites is $y=1-4p$.  (Note that $p \leq 1/8$, so that $y \geq 1/2$.)  Thus we have the
equation of motion $\dot{p} = z(1-7p)^7/(1-4p)^6 - p$.

To study the phase transition, the approximation developed above needs to be refined
somewhat.  First, we introduce bond fractions $q_i$ and $p_i$ and site fractions $y_i$
for each sublattice, $i = 1,...,8$.  In a maximally occupied configuration all the bonds
in one sublattice are occupied; thus $p_i \leq 1$.  (Of course, if $p_i = 1$ for one sublattice, it should
be zero for the others.)  A convenient labeling scheme for the sublattices is shown
in Fig. 13.  Now, for a vacant bond in sublattice $j$ to be available for insertion,
the six neighboring bonds, all of which belong to other sublattices, must be vacant as well.

Let $r_j$ be the number of {\it half-occupied} bonds in sublattice $j$.  Each occupied bond
in one of the neighboring sublattices (i.e., having a shared vertex) contributes to $r_j$;
for example,

\begin{equation}
r_1 = p_2 + p_4 + p_5 + p_6 + p_7 + p_8.
\end{equation}

\noindent Knowing the $r_j$ and $p_j$, we have for the empty bond fractions,
$q_j = 1 - p_j - r_j$, while $y_j = 1 - p_j - r_j/2$.
The fraction of occupied bonds in sublattice 1 follows

\begin{equation}
\frac{dp_1}{dt} = z \frac{q_1 \, q_2 \, q_4 \, q_5 \, q_6 \, q_7 \, q_8}{y_1^6 } - p_1
\label{dp1dt}
\end{equation}

\noindent with analogous equations for the other sublattices.
The set of eight equations for the occupation fractions are
integrated at fixed $z$, using a fourth-order Runge-Kutta scheme,
until a steady state is reached.  The presence of order is indicated
by a nonzero value of $\phi = p_{max} - p_{min}$, i.e., of the difference
between the largest and smallest sublattice coverages.  Starting from an
ordered initial distribution ($\phi_0 \geq 0.7$), the final state is disordered
for $\mu < \mu_< = 1.8378$, but takes a nonzero value ($\phi \geq 0.8$), for $\mu > \mu_<$.
The density jumps from $\rho=0.179$ to 0.207
at the transition.
The transition point $\mu_<$ should interpreted as
a stability limit of the ordered phase (i.e., as a spinodal) and not
as the coexistence value.  Indeed, the transition occurs at
higher $\mu$ values for smaller initial values of $\phi$; for $\phi_0 = 0.6$,
for example, it happens at $\mu = 1.9023$.
The coexistence density and chemical potential are not furnished by
the present dynamic approximation, since it
does not provide an estimate of the free energy.  The dynamic pair approximation
nevertheless captures the discontinuous nature of the phase transition.

\begin{figure}[!htb]
\includegraphics[clip,angle=0,width=0.8\hsize]{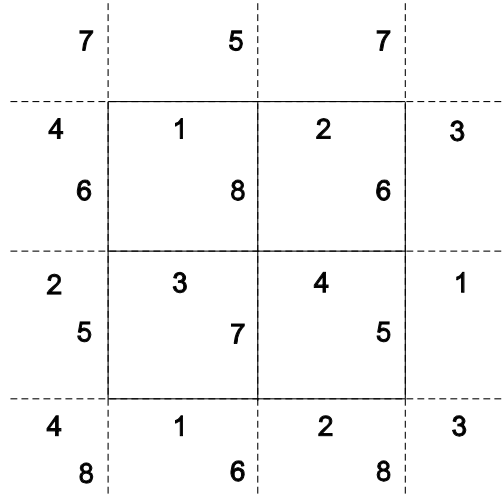}
\vspace{-4em}

\caption{Sublattice labeling scheme used in the pair approximation.
}
\label{palabels}
\end{figure}

\bibliographystyle{apsrev}

\end{document}